\documentclass[a4paper]{llncs}

\usepackage{amssymb}
\usepackage{graphicx}
\usepackage{subfigure}
\usepackage{floatflt}
\usepackage{listings}
\usepackage{multicol}

\usepackage{rotating}

\usepackage{url}
\urldef{\mailsa}\path|daniel.ritter@sap.com|
\newcommand{\keywords}[1]{\par\addvspace\baselineskip
\noindent\keywordname\enspace\ignorespaces#1}

\begin{document}

\mainmatter  

\title{A Logic Programming Approach to Integration Network Inference}


%
%
\author{Daniel Ritter}
%

\institute{SAP AG, Technology Development -- Process and Network Integration,\\
Dietmar-Hopp-Allee 16, 69190 Walldorf, Germany\\
\mailsa\\
}

\maketitle

\begin{abstract}
The discovery, representation and reconstruction of (technical) integration networks from Network Mining (NM) raw data is a difficult problem for enterprises. This is due to large and complex IT landscapes within and across enterprise boundaries, heterogeneous technology stacks, and fragmented data. To remain competitive, visibility into the enterprise and partner IT networks on different, interrelated abstraction levels is desirable.

We present an approach to represent and reconstruct the integration networks from NM raw data using logic programming based on first-order logic. The raw data expressed as integration network model is represented as facts, on which rules are applied to reconstruct the network. We have built a system that is used to apply this approach to real-world enterprise landscapes and we report on our experience with this system.

\keywords{Datalog, Integration Networks, Knowledge Representation, Logic Programming, Network Inference, Network Mining}
\end{abstract}

\section{Introduction}
Enterprises are highly connected to partners and even competitors as part of value chains consisting of business processes. The business document exchange is actually implemented by complex, underlying networks of application and middleware systems, called integration networks. To remain competitive enterprises have to adapt their business processes in a timely and flexible manner, which requires visibility and control over the integration network. However, currently information is locked into systems of an enterprise. To overcome this situation, a new discipline, called Network Mining (NM), strives to discover and extract raw data hidden within heterogeneous systems in complex enterprise landscapes \cite{confenis2012,lsna2012}. The raw data implicitly contains information about the integration network, i.e. middleware and application. From that, our system reconstructs integration networks. For the system user, the resulting linked real-world data describing the "as-is" network can then be captured in e.g. network-centric BPMN models \cite{bpmn2011}.

\begin{figure*}
\centering
\includegraphics[width=\textwidth]{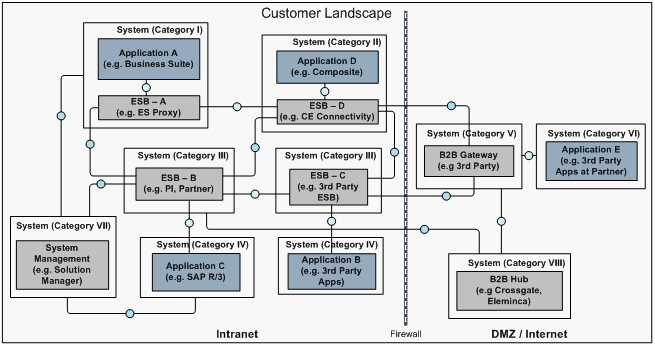}
\vspace*{-.5cm}
\caption{Sample (technical) Integration Network showing logical systems as participants with embedded integration capabilities and standalone middlewares as well as B2B gateways}
\label{fig:integrationNetwork_all}
\end{figure*}

A generalized view of such a network is shown in Fig. \ref{fig:integrationNetwork_all}. When looking at an enterprise landscape, the systems within the integration network can be classified into different categories based on the integration content and the role they play. The classification provides insight into the capabilities and complexity of the network and allows to manage business processes, contextualized visualization and operation on the network. These categories span from applications with embedded integration or even mediation capabilities, like proxies, enterprise services, composite applications or applications with service adaptation (Categories I+II), over standalone Enterprise Service Bus (ESB) or middleware instances with flexible pipeline processing, e.g. mapping, routing and connectivity for legacy systems (Category III+IV), to Business to Business (B2B) gateways for cross-enterprise document exchange (Categories V+VI) and system management solutions, which allow to operate these systems, their software and lifecycle (Category VII).
\vspace*{-.5cm}
\vspace*{-.5cm}

In this paper we present an approach to model and reconstruct integration networks from discovered raw data using logic programming, more precisely standard Datalog with recursion and stratified negation. We describe how information in form of NM raw data can be represented independent of their original domain in a Network Integration Model (NIM) and how user facts can be added. We have chosen Datalog to represent this model, which we use to develop Datalog programs (i.e. a finite set of Datalog rules) that express the network. That means identifying entity equivalences, computing edges and semantic references as well as dealing with user input. We validated our approach on simulated integration network data and report our experience with the network inference Datalog system in real-world enterprise networks as well as possible extensions.

In Section \ref{sec:motivation} we describe the problem domain and state on design principles and decisions in Section \ref{sec:design}. Section \ref{sec:naim} defines the NIM and Section \ref{sec:inia} introduces the inference algorithm. Section \ref{sec:results} shows experimental results and states on experiences. Section \ref{sec:relatedWork} concludes with related work, before we draw conclusions and outline future research in Section \ref{sec:outlook}.

\section{Motivation} \label{sec:motivation}
Our premise is that relevant data for computing integration networks is hidden in enterprise system landscapes. However, for that it has to be discovered by NM from mostly disjoint domains in different formats with different meaning \cite{lsna2012}. The integration networks derived from the discovered information consist of nodes and edges on different abstraction levels.

\begin{figure*}[!ht]
\begin{center}$
\begin{array}{cc}
\subfigure[Schematic view on NM raw data in common network representation]{\label{fig:challenges_1}\includegraphics[width=0.5\textwidth]{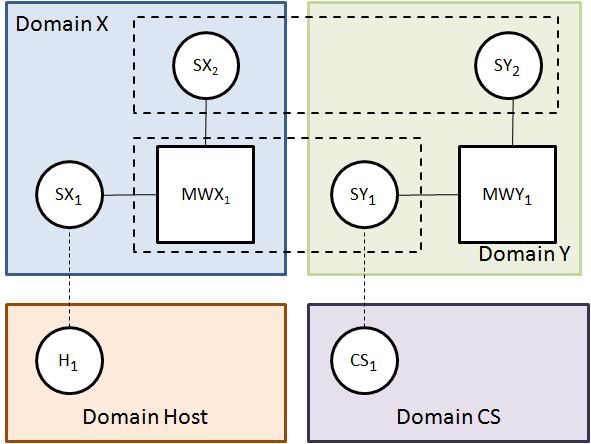}}
\subfigure[Schematic view of reconstructed network]{\label{fig:challenges_2}\includegraphics[width=0.5\textwidth]{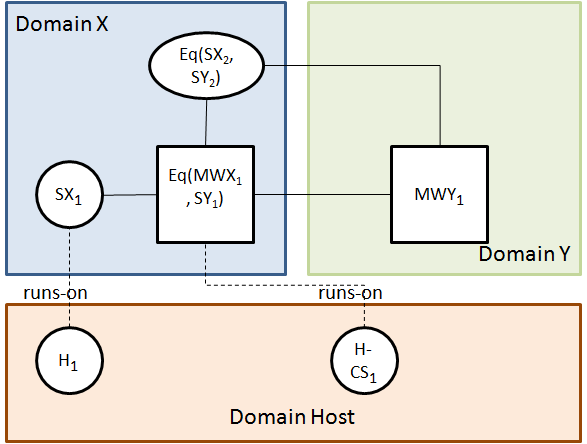}} &
\end{array}$
\end{center}
\vspace*{-.5cm}
\caption{Schematic view on the inference challenges}
\label{fig:challenges}
\end{figure*}

The basic entities of the integration network are logical systems (e.g. tenants, applications, integration middleware) and message flows, which are either direct connectivity or mediated communication/ integration. The actual information about these entities as well as their semantics are discovered by Network Mining (NM) systems \cite{lsna2012}. However, the discovered raw data is domain-specific and needs to be translated into a domain independent model for network inference, while preserving its semantics. The definition of a Network Integration Model (NIM) is the basis for applying network inference algorithms. Since the raw data comes from disjoint domains, in different formats with different semantics, inference algorithms have to deal with possibly duplicate, fragmented, uncertain or incorrect information while computing the network. Fig. \ref{fig:challenges} schematically shows some of these challenges. For instance, entity equivalences have to be identified and handled. Direct and transitive edges have to be calculated and semantic relations between nodes have to be inferred. Fig. \ref{fig:challenges_1} shows systems $SX_1$ and $SX_2$ discovered from domain $X$ exchanging messages over middleware system $MWX_1$, and systems $SY_1$ and $SY_2$ discovered from domain $Y$ exchanging messages over middleware system $MWY_1$. Here, $SX_2$ and $SY_2$ denote the same system, as well as $MWX_1$ and $SY_1$ are equivalent. Based on the inferred equivalences, the nodes are partitioned as equivalence relations $Eq$, i.e. $Eq(MWX_1, SY_1)$ and $Eq(SX_2, SY_2)$, and the edges are computed accordingly (see Fig. \ref{fig:challenges_2}). Systems or applications run on physical hosts, e.g. $H_1$ from discovery domain $Host$. The relationships between systems and hosts are not considered as edges but semantic references within the network. Hosts build the bridge to the related domain of system management networks, which are addressed by \cite{itsm2006,xen2009}. A new host $CS_1$ is added to the network as user knowledge on which $SY_1$ runs. When merging systems $MWX_1$ and $SY_1$ the semantic relation is preserved.

\section{Design Principles and Decisions} \label{sec:design}
The major design decisions taken were about finding a representation for an integration model and a language to express inference algorithms. We needed to select (1) an approach, which does not require to modify the system when changing the inference programs or the integration model, (2) a well-understood representation for information suitable for the inference approach, and (3) a sufficiently powerful inference technique, simple enough to be used by our customers and partners to define their own inference programs. 

The necessity of (1) is derived from developing the inference programs in the early prototypes. The domain of the data and the scope of inference evolved - and it will continue to do so as more data sources are integrated and inference is refined. Hence the lifecycle of the data model and of the inference programs needs to be decoupled from that of the system. Since system landscapes and business networks for large enterprises are very complex and many implementations need customer-specific modifications or extensions both (2) and (3) are required. As the relational model is a foundation for most business applications and is thus well-understood by customers, it is a natural choice for (2). Consequently, we initially considered SQL and its imperative extensions to express inference programs.
However, as network analysis and inference are expressed more naturally using recursive rules we moved towards logic programming languages like Prolog or Datalog, choosing Datalog for its simpler semantics.

\section{The Integration Network (Inference) Model} \label{sec:naim}
The model for representing integration networks as virtual "as-is" enterprise landscape covers a representative intersection of entities from the enterprise integration middleware space \cite{eip}. Although this domain has many aspects, which are even differently treated in different system implementations, we identified a common, core meta-model, which we call Network Integration Model (NIM). The basic NIM entities relevant for the inference are introduced subsequently, while more entities might be explained later where necessary.

\begin{figure*}[!ht]
\centering
\includegraphics[width=0.75\textwidth]{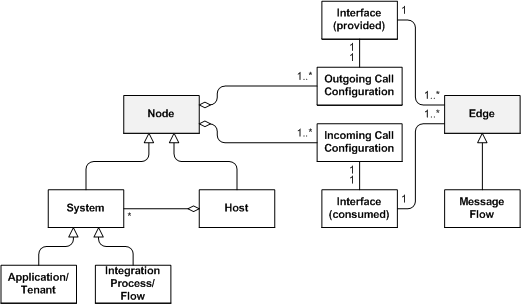}
\vspace*{-.3cm}
\caption{The basic NIM entities and their relations}
\label{fig:sample_business_trading_network}
\end{figure*}

The base premise for defining an integration meta-model is to represent the actual physical hosts in the enterprise landscape as first class entities and then find the interfaces provided or called by them during message-based communication. Since most of the communication actually addresses logical entities like applications or tenants, called systems, running on the physical hosts, a \emph{System} is considered a node of the network. That means, systems represent (business) application and integration logic. For the communication with other systems via messages the \emph{MessageFlow} represents edges in the network. Technically, messages are exchanged over interfaces, \emph{Interface}, and channels, containing e.g. service bindings and operations, which we represent as \emph{IncomingConfiguration} and \emph{OutgoingConfiguration}. The inbound and outbound configurations are considered separate entities, since they carry important information about the message flows, thus helping to reconstruct the network's edges. This notion can also be found in a common graph traversal algebra to set custom processors or actions when entering or leaving a node \cite{gdm2011}. Fig. \ref{fig:sample_business_trading_network} shows the basic NIM entities and their relations.

\section{The Network Inference Approach} \label{sec:inia}
The algorithm for computing integration networks consists of multiple steps, which have been identified for a parallel analysis allowing it to scale across large datasets of NM raw data. Since the information is represented in the NIM, the inference mechanism is independent of the specific integration and system domains. As discussed in Section \ref{sec:motivation}, unique systems and hosts are identified by equivalence algorithms and semantic links between hosts and systems are computed (step 1). Based on that, incoming and outgoing configurations are identified (step 2) and then used to reconstruct message flows through building separate call graphs (steps 3,4) which are merged afterwards (step 5). Then message flows are linked with application and integration content (step 6) and user knowledge is integrated. With user knowledge, the quality of the inference mechanism can be improved and information complemented or enriched. Within the inference programs, all user knowledge literals end with the "user" postfix, while discovered knowledge ends with "disc" (i.e. edb relation). 

To formalize the network reconstruction, a logic programming approach is used, in which the algorithms are described by Datalog rules and the discovered raw data is a set of Datalog facts according to NIM. The different processes of adding newly discovered information and removing outdated is continuous. For that, each piece of discovered information is annotated with a timestamp. However, instead of removing outdated information that is referenced by higher layer information models as in \cite{bpmn2011}, it is kept and marked outdated until it is not referenced anymore.


\emph{Step 1: Identify unique hosts and systems} To identify hosts and systems uniquely through building equivalence classes, the single instances have to be identified. While hosts can be identified by e.g. host name, IP-address, the systems have no universally applicable identification scheme, thus they are usually identified using context dependent identifiers. For instance, the set of host identifiers can be an IP-address, the DNS name, and a host name. This information mainly comes from different, disjoint instances of system management software, mostly from IT service management \cite{itsm2006} and virtualization systems \cite{xen2009}. All identifiers are contained in the equivalence class and any reference to one of them identifies the host. While these equivalence classes are not stable over time, it is quite likely that at least one of the elements of an equivalence class does not change if another one changes, thus making the identification more robust. That way, identity can be maintained over long periods of time in the presence of constant but gradual change. The raw facts from NM are $host\_disc(host\_id, URI)$ and $system\_disc(sys\_id, URI)$, which relate a $host\_id$ or $sys\_id$ to an addressable URI. Relations like $same\_host\_disc(host\_id1, host\_id2)$ and $same\_sys\_disc(sys\_id1, sys\_id2)$ connect two host or system identifiers, e.g. which refer to the same physical host or logical system. The semantic relation \\$runs\_on\_disc(sys\_id, host\_id)$ connects a system to the host that it runs on. For simplicity, homogenous clusters of machines are also considered as one host.

\begin{lstlisting}[basicstyle=\slshape, breaklines=true, mathescape, frame=single,numbers=none, caption={Host equivalence exploiting information about system landscape}, label={listing:samehost}]
same_sys(?sys_id1, ?sys_id2) :- 
    same_sys_disc(?sys_id1, ?sys_id2).  
same_sys(?sys_id1, ?sys_id2) :- 
    same_sys(?sys_id1, ?sys_id3), 
    same_sys(?sys_id3, ?sys_id2). 

same_host(?host_id1, ?host_id2) :- 
    runs_on_disc(?sys_id1, ?host_id1), 
    runs_on_disc(?sys_id2, ?host_id2), 
    same_sys(?sys_id1, ?sys_id2).
\end{lstlisting} Based on that, rules for e.g. $same\_sys$ and $same\_host$ are used to infer equivalence classes that allow to write rules that exploit the information about system landscapes. For instance, more than one system can run on one physical host, but one system cannot run on more than one host, Listing \ref{listing:samehost}.

\begin{figure*}[!ht]
\begin{center}$
\begin{array}{cc}
\subfigure[Outgoing/ incoming configuration call graph]{\label{fig:outincallgraph}\includegraphics[width=0.35\textwidth]{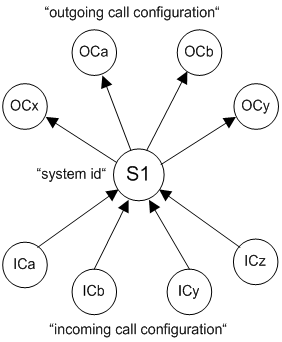}}
\subfigure[Call graph extension]{\label{fig:extcallgraph}\includegraphics[width=0.65\textwidth]{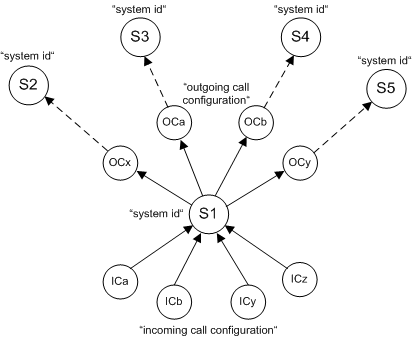}} &
\end{array}$
\end{center}
\vspace*{-.3cm}
\caption{Outgoing and incoming configuration call graphs}
\label{fig:businessNetwork_all}
\end{figure*}

\emph{Step 2: Determine Incoming and Outgoing Calls} In current middleware route configurations, the senders of incoming calls to the system can be registered but are mostly unknown. On the other hand, components like the file adapter and the Apple Push Notification Service (APNS) always contain the sender system \cite{eip}. However, for outgoing calls from the sender system, e.g. via HTTP, SOAP, receiver or outgoing call configurations are needed to initiate the message flow to the receiver. This results in an outgoing and incoming call graph depicted in Fig. \ref{fig:outincallgraph}. The $incoming\_disc(sys\_id, URI)$ and $outgoing\_disc(sys\_id, URI)$ facts relate a \emph{sys\_id} to a $URI$ of an incoming configuration or an outgoing configuration for the identified system.
\vspace*{-.5cm}

\emph{Step 3: Determine Message Flows based on Outgoing Calls}  Since outgoing calls are made to a particular endpoint, the corresponding call configurations contain an identifier for the receiving host or system. These identifiers can then be matched against the identifiers that were determined in step 1. If no identifiers are available, these call configurations are processed in step 4. To relate outgoing call configurations to receiver systems $recv\_disc(URI, sys\_id)$ relates a $URI$ to an outgoing configuration to a $sys\_id$ that identifies a receiving system or similarily $recv\_host\_disc(URI, host\_id)$ for hosts.
\begin{lstlisting}[basicstyle=\slshape, breaklines=true, mathescape, frame=single,numbers=none, caption={Message flow from outgoing configuration}, label={listing:messageflow}]
msg_flow(?sys_id_snd, ?sys_id_recv) :- 
    outgoing_disc(?sys_id_snd, ?RCONF), 
    recv_disc(?RCONF, ?sys_id_recv).
\end{lstlisting}


\begin{lstlisting}[basicstyle=\slshape, breaklines=true, mathescape, frame=single,numbers=none, caption={Message flow for host configurations}, label={listing:messageflowhost}]
msg_flow_host(?host_id_send, ?host_id_recv) :- 
    runs_on_disc(?sys_id_snd, ?host_id_send), 
    outgoing_disc(?sys_id_snd, ?RCONF), 
    recv_host_disc(?RCONF, ?host_id_recv).
\end{lstlisting}
Then $message\_flow(sys\_id\_snd, sys\_id\_recv)$ rules determine the message flows between systems (Listing \ref{listing:messageflow}) and \\$message\_flow\_host(host\_id\_send, host\_id\_recv)$ between hosts (Listing \ref{listing:messageflowhost}). That results into a an extension of the call graph shown in Fig. \ref{fig:extcallgraph}, in which $S1$ represents a system connected to other systems via incoming and outgoing configurations.

\emph{Step 4: Determine Message Flows based on Incoming Calls} Similar to the previous step, incoming call configurations are identified. For that, $send\_disc(URI,$ \\ $sys\_id)$ facts are related via URI to incoming configurations. Again, this results in an extension of the call graph.

\begin{figure*}[!ht]
\centering
\includegraphics[width=3.5in]{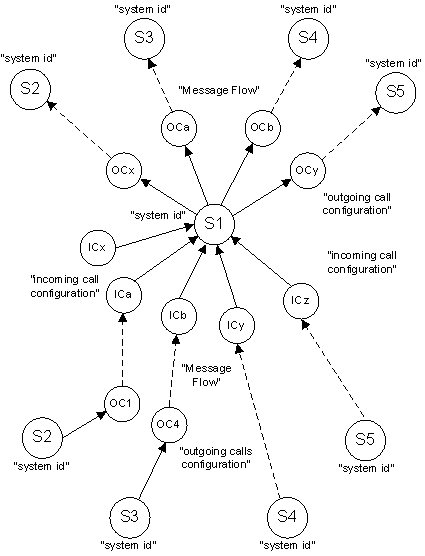}
\caption{Call graph extended by merged information from different systems}
\label{fig:mergedgraph}
\end{figure*}

\emph{Step 5: Merge Call Graphs for a System} So far unique hosts and systems are identified and message flows are determined for a single system. Now, the identified incoming and outgoing call configurations from different systems are matched. This is done by matching compatible protocols, message types, etc. After new message flows are identified, the call graph is extended by the merged information (see Fig. \ref{fig:mergedgraph}). In case some incoming or outgoing call configurations do not match to already identified call configurations, they are kept in the model as "unlinked" configurations for matching new configurations.

\emph{Step 6: Link Message Flows to Application and Integration Content} The outgoing and incoming call configurations with hosts and systems result in a view of the network. However, these message flows only conclude communication between hosts and systems. The outgoing and incoming call configurations also have a link to application and integration content deployed and running on the systems. This content refers to the particular process or integration steps that trigger outgoing calls or receive incoming calls. In other words, process models \cite{processMining2011} and middleware routes \cite{eip}, i.e. integration flow (IFlow) or integration process, give insight into the details within systems and hosts and could be used to correlate operational data to trace messages through middleware systems. 

\begin{lstlisting}[basicstyle=\slshape, breaklines=true, mathescape, frame=single,numbers=none, caption={Identifying IFlows}, label={listing:integrationflow}]
iflow(?sys_id_snd, ?sys_id_recv, ?sys_id_mw, ?URI) :- 
    msg_flow_disc(?sys_id_snd, ?sys_id_mw, ?URI), 
    msg_flow_disc(?sys_id_mw, ?sys_id_recv, ?URI).
\end{lstlisting}

For instance, the IFlow $iflow(sendsys\_id, recvsys\_id, mwsys\_id, URI)$ relates senders to receivers through a middleware system, which can be calculated e.g. through the rule in Listing \ref{listing:integrationflow}. 


\section{Results and Experiences} \label{sec:results}
For the evaluation of our approach, we used our Datalog system, which is a basic Datalog implementation in Java/OSGi based on \cite{datalog}, that allows to evaluate recursive rules and supports basic data types, comparisons and expressions in Datalog rules. The raw data comes from our Network Mining prototype, which discovers information in our testbed and transforms it to NIM Datalog facts. The testbed consists of two middleware systems, i.e. \emph{HXP} and \emph{H73}, of different releases for mediated communication, and have embedded IDoc and WebService capabilities for direct communication and a System Landscape Directory (SLD) for system management information. This setup contains real-world conditions which we found in our customer landscapes, e.g. cross-middleware inference, combination of embedded and mediated communication, fragmented information registered in different domains.

The results of the experiments are shown, e.g. for systems and message flows in \emph{HXP} in Table \ref{tab:results_hxp_nodes}, \ref{tab:results_hxp_edges} and for \emph{H73} in Table \ref{tab:results_h73_nodes}, \ref{tab:results_h73_edges}. The tables show two aspects of the system, namely the discovery and the inference quality. For the inference, the entries for systems and message flows as well as top-level connections are important. The discovery is mainly depicted by attribute entries for the network entities and show minor gaps in the discovery process., e.g. in the category "Correct System Attributes" (see Table \ref{tab:results_hxp_nodes}).
\begin{table*}[h!b!p!]
\caption{Inference Result on NM Testbed (Nodes of HXP-PI System)}
\label{tab:results_hxp_nodes}
\begin{tabular}{lll}
\hline
Category & Absolute Value & Percentage \\
\hline
\emph{Found expected Systems} & 12 & 100\% \\
Correct System Attributes & 35 & 64\% \\
System Attributes with Limitation & 20 & 36\% \\
\hline
\end{tabular}
\end{table*}
For the HXP-PI system, $12$ nodes and $55$ node attributes are expected (see Table \ref{tab:results_hxp_nodes}). In total $13$ top-level connections are expected which group $31$ message flows (see Table \ref{tab:results_hxp_edges}). Furthermore, the top-level connection have $26$ attributes, while the message flows have $372$ attributes.
\begin{table*}[h!b!p!]
\caption{Inference Result on NM Testbed (Edges of HXP-PI System)}
\label{tab:results_hxp_edges}
\begin{tabular}{lll}
\hline
Category & Absolute Value & Percentage \\
\hline
\emph{Found Expected Top-Level Connection Groups} & 13 & 100\% \\
Correct Top-Level Connection Group Attributes & 26 & 100\% \\
\\
\emph{Found Expected MessageFlows} & 31 & 100\% \\
Correct MessageFlow Attributes & 337 & 91\% \\
MessageFlow Attributes with Limitation & 34 & 9\% \\
\hline
\end{tabular}
\end{table*}

For the cross middleware systems and message flow inference, in total $18$ unique, logical systems were inferred from $29$ partially duplicate raw facts via equivalence determination (see Table \ref{tab:results_hxp_nodes} and \ref{tab:results_h73_nodes}) and $34$ message flows have to be reconstructed and grouped to $17$ top-level connections using incoming and outgoing call graph merge operations. For instance, logical system \emph{HXP\_105} from \emph{PI-HXP} with runs-on host id \emph{xxx2474} from SLD was found in the middleware configuration and SLD system information facts and merged into an equivalence class (see Table \ref{tab:results_detail_nodes}). At the same time, the corresponding message flows between \emph{HXP\_105} to \emph{HXP\_106} were reconstructed from PI configuration (conf.) and runtime (runt.) data connected to the system equivalence sets and merged into an top-level connection group (see Table \ref{tab:results_detail_edges}). The group consists of the message flow over sender interface \emph{FlightSeatAvailQuery} to system \emph{HXP\_106}, which checks for free seats and is followed by a message to the same system over interface \emph{BookOrderRequest} in case of a positive answer to the first query. If the booking order request was successful, system \emph{HXP\_106} answers over interface \emph{FlightBookOrderConfirm} to confirm the request. No unexpected systems or message flows were found and the complete network structure was reconstructed correctly.

Similarly, the H73-PI system has 3 parties, i.e. B2B contexts, 6 expected nodes with 31 attributes (for Table \ref{tab:results_h73_nodes}), 4 top-level connections, grouping 6 flows (for Table \ref{tab:results_h73_edges}), with 8 attributes on the top-level connections and 78 on the message flows. 
\begin{table*}[h!b!p!]
\caption{Inference Result on NM Testbed (Nodes of H73-PI System)}
\label{tab:results_h73_nodes}
\begin{tabular}{lll}
\hline
Category & Absolute Value & Percentage \\
\hline
\emph{Found expected Parties} & 3 & 100\% \\
Found expected System & 6 & 100\% \\
Correct System Attributes & 22 & 71\% \\
\hline
\end{tabular}
\end{table*}

\begin{table*}[h!b!p!]
\caption{Inference Result on NM Testbed (Edges of H73-PI System)}
\label{tab:results_h73_edges}
\begin{tabular}{lll}
\hline
Category & Absolute Value & Percentage \\
\hline
\emph{Found Expected Top-Level Connection Groups} & 4 & 100\% \\
Correct Top-Level Connection Group Attributes & 8 & 100\% \\
Top-Level Connection Group Attributes with Limitation & 0 & 0\% \\
\\		
\emph{Found Expected MessageFlows} & 6 & 100\% \\
Correct MessageFlow Attributes & 75 & 96\% \\
MessageFlow Attributes with Limitation & 3 & 4\% \\
\hline
\end{tabular}
\end{table*}

The detailed inference results are only shown partially due to the mass of data discovered. Hence Table \ref{tab:results_detail_nodes} shows an excerpt of the results of systems with the discovered description, the inferred host and the equivalence class denoted by "discovered system". Similarly, an excerpt of the inferred message flows are shown in Table \ref{tab:results_detail_edges}. For that, the top-level connections, i.e. grouped message flows are listed with their message flows denoted by sender and receiver and the type of discovered facts from which the data came from (as "From"). In the excerpt, all message flows themselves build an equivalence class of same flows found in runtime logs (runt.) and configuration (config.).

\begin{table*}[h!b!p!]
\caption{Excerpt of HXP-PI system Inference Result}
\label{tab:results_detail_nodes}
\begin{tabular}{lllll}
\hline
System (Name)&Description&Host&Discovered System \\
\hline
HXP\_105&Booking System&HXP on xxx2474&PI as Bus. System\\
&&SLD as Bus. System\\
\hline
HXP\_106&Lufthansa&HXP on xxx2474& PI as Bus. System\\
&&SLD as Bus. System\\
\hline
HXP\_107&American Airlines&HXP on xxx2474&PI as Bus. System\\
&&SLD as Bus. System\\
\hline
...&...&...&...\\
\hline
Interflug&Interflug&unknown&PI as Bus. Component\\
\hline
...&...&...&...\\
\hline
Singapore&Singapore Airlines&unknown&PI as Bus. Component\\
\hline
\end{tabular}
\end{table*}

\begin{table*}[h!b!p!]
\caption{Excerpt of HXP-PI message flow and top-level Inference Result}
\label{tab:results_detail_edges}
\begin{tabular}{llllll}
\hline
Top-level Connect. \\Group&Interface&Sender&Receiver&From\\
\hline
HXP\_105$<->$HXP\_106&BookOrderRequest&HXP\_105&HXP\_106&Config.+Runt.\\
&FlightSeatAvailQuery&HXP\_105&HXP\_106&Config.+Runt.\\
&FlightBookOrderConfirm&HXP\_106&HXP\_105&Config.+Runt.\\
\hline
...&...&...&...&...\\
\hline
\end{tabular}
\end{table*}

\begin{figure*}[!ht]
\centering
\includegraphics[width=\textwidth]{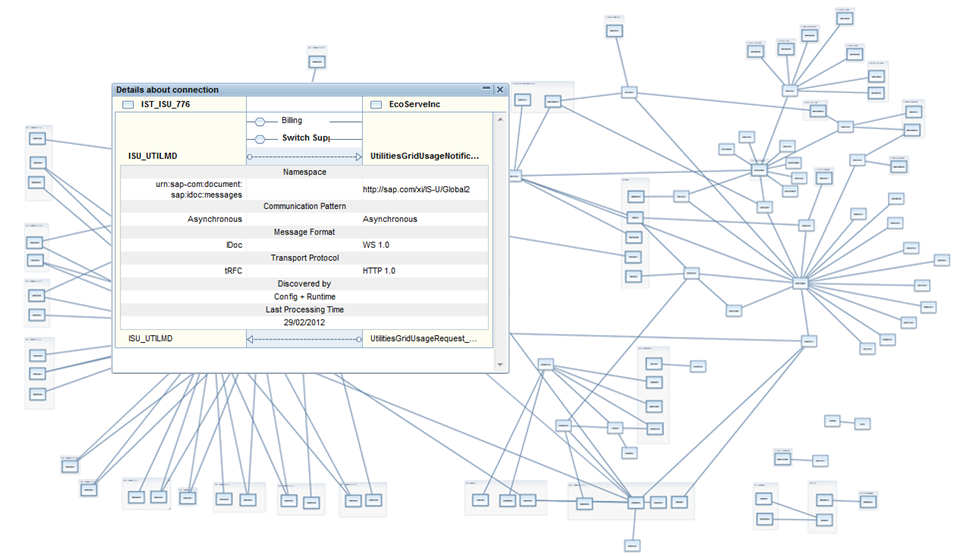}
\caption{Real-world customer network in a network-centric BPMN notation \cite{bpmn2011} inferred from NM raw data showing network structure and detailed view for one edge}
\label{fig:sample_landscape}
\end{figure*}

Due to good results in our testbed, we applied the system to real-world customer landscapes as shown in Fig. \ref{fig:sample_landscape}. This real-world validation was very successful on both counts. Firstly, it proved that the auto-discovery and inference is indeed feasible and resulted in highly reliable results. Secondly, our system would be quite helpful in the everyday work of an integration architect, consultant or integration developer, since it gives an overview of the complete integration network which is currently not possible within the integration middleware tools. The system reduces the effort to document integration scenarios substantially, in particularly by a foreseen export of network details into PDF or office format. That helps to answer specific questions about the network, which are currently still impossible (or difficult) to achieve. For example, when combining configuration and runtime data it is possible to find connections that are not used any longer or were seldom used in a given period of time. Hence, one of the customers plan an upgrade project and with such a system a substantial migration time and effort will be saved.
\vspace*{-.5cm}

\section{Related Work} \label{sec:relatedWork}
Our approach for integration network represention and inference is based on Datalog, which is a well-researched topic \cite{datalog1978,datalog} that had its revival recently due to good parallelization capabilities, latest through the work of Hellerstein et al. \cite{boom2010,declarativeImperative}. Even in the enterprise analytics domain, Datalog was recently applied, mainly through work of \cite{logicblox2010,logicblox2011,logicblox2011_2}. However, these approaches address non-network inference domains for which they define extensions.

In terms of the meta-model for integration network, \cite{gdm2011} represents closest known related work, in which a path algebra is defined that is used to traverse arbitrary graphs. Similarly we define nodes and edges with inbound and outbound connectors, however different in terms of meaning and usage.

For NM systems in general, related work is conducted in the area of Process Mining (PM) initiated by \cite{processMining2011}, which sits between computational intelligence and data mining. It has similar requirements for data discovery, conformance and enhancement with respect to NM \cite{lsna2012}, but does not work with network models and inference. PM exclusively strives to derive BPM models from process logs. Hence PM complements NM in the area of business process discovery.

Gaining insight into the network of physical and virtual nodes within enterprises is only addressed by the \emph{Host} entity in NIM, since it is not primarily relevant for visualizing and operating integration networks. This domain is mainly addressed by the IT service management \cite{itsm2006} and virtualization community \cite{xen2009}, which could be considered when introducing physical entities to our meta-model.

The linked (web) data research shares similar approaches and methodologies, which have so far neglected linked data within enterprises and mainly focused on RDF-based approaches \cite{linkedData2009,LinkedWebData2009}. Applications of Datalog in the area of linked data \cite{datalogWeb2010,datainweb2010} and semantic web \cite{semanticweb2010} show that it is used in the inference domain, however not used for network inference.

\section{Discussion and Future Work} \label{sec:outlook}
In this paper we introduce a new domain for information discovery, machine learning, and network reconstruction, for which we defined a modeling and inference approach to reconstruct integration networks from NM raw data using Datalog.
The network model developed specifically for the connectivity and integration domains and covers an intersection of the relevant entities, which we derived through the analysis of several middleware systems on the market. We encoded the discovered raw data as Datalog facts to create a domain independent knowledge base and applied rule-based inference representing a multi-step network inference approach. 
We validated our approach on a simulated integration network and reported our experiments on applying our system to real-world enterprise networks. The evaluation shows good results with respect to the challenges like equivalence class determination, flow- and cross-middleware network reconstruction as introduced in Section \ref{sec:motivation}. Although the network structure could be reconstructed very well, the discovery range should be improved to attach more integration details to the attributes of the network entity instances.

Future work will be conducted in several areas, among them the improvement of the discovery range, the inference of business process models from NM data and the correlation to integration networks as well as extensions to standard Datalog to improve the current implementation. For instance, the efficient compilation of Datalog programs to current hardware \cite{neumann2011}, distributed systems \cite{hadoop2012} or pruning with CHR \cite{chr2011} could guarantee more efficient Datalog processing. Since not all facts have the same certainty, we will also look into probabilistic extensions of Datalog like \cite{constraints2010,constraints2011}, which could help to express different levels of certainty with respect to network model instances. The work conducted in \cite{dedalus2011} will be considered for time aspects, which could help to prune large, outdated networks from system landscapes with historical data.



\end{document}